\documentclass[preprint,preprintnumbers,amsmath,amssymb]{revtex4}

\begin{document}

\title{Equivalence of different descriptions for $\eta$ Particle in Simplest Little Higgs Model}

\author{Ran Lu, Qing Wang}

\address{Department of Physics,Tsinghua University,Beijing 100084,P.R.China\footnote{mailing address} \\
    Center for High Energy Physics, Tsinghua University, Beijing 100084, P.R.China}

 \email{luran@mails.tsinghua.edu.cn}
\date{\today}

\begin{abstract}
In $SU(3)$ simplest little Higgs model, a characteristic particle is
the light pseudoscalar boson $\eta$, it leads interesting signals in
the LHC/ILC and is studied in literature with different
parameterizations. In this work, we show that these different
descriptions for $\eta$ particle are equivalent up to some $SU(3)$
rotations as long as we suitably redefine the pseudo Goldstone boson
fields. We evaluate the necessary $SU(3)$ rotations and built up
explicit expressions for redefined fields.

\end{abstract}
\maketitle


The Higgs boson is the last ingredient of the Standard model (SM),
its properties attract lots of our attentions either in experiments
and in theories. Experimentally, precision measurements of
electroweak parameters give indirect limit on the $m_H$ to be less
than 186 GeV at 95$\%$ confidence level(C.L.)\cite{mHupperBound} and
direct search by LEPII experiments gives a lower bound of 114.4GeV
at the 95$\%$ C.L.\cite{mHlowerBound}. Theoretically, numerous new
physics models are invented with different roles of Higgs in the
models. Among them, a class of little Higgs models are proposed to
soften hierarchy problem \cite{littleHierarchy}. In these models, relatively
light Higgs boson mass is due to its identity as a pseudo Goldstone
boson of some enlarged global symmetries. Among various little Higgs
models, the simplest little Higgs model \cite{simplest,simplest1} is
attractive due to its relative simple theory structures. The model
is based on the $[SU(3)\times U(1)_X]^2$ global symmetry with its
diagonal subgroup $SU(3)\times U(1)_X$ gauged. The vacuum
expectation values of two $SU(3)$ triplet scalar fields $\Phi_1$ and
$\Phi_2$ spontaneously break both the global and the gauge
symmetries. Uneaten pseudo Goldstone bosons are $SU(2)_L$ doublet
$h$ and a pseudoscalar $\eta$. The typical feature of simplest
little Higgs model is the existence of this $\eta$ particle, as a
pseudo Goldstone boson, it should be light and the upper and lower
bounds of its mass were estimated in Ref.\cite{axion}, roughly
speaking in GeV range below electroweak scale $v=$250GeV. Due to its
lightness, $\eta$ may affects the physics significantly and leads
interesting phenomenologies at next generation of high energy
colliders \cite{axion,axion1,decay,neutrino,neutrino1}. For the
theoretical computations on the phenomenologies related to $\eta$
particle, pseudo Goldstone bosons $h$ and $\eta$ should be written
as a combination field $\Theta$ which is defined as
\begin{equation}
        \Theta =
        \frac{1}{f}[\frac{\eta}{\sqrt{2}}T+\Theta_h]\hspace{1cm}\Theta_h=
        \begin{pmatrix}
                0 & h \\
                h^\dagger & 0 \\
        \end{pmatrix}\label{ThetaDef}
\end{equation}
$h$ is a doublet which will become the SM Higgs, $\eta$ is a real
pseudoscalar field, $T$ is its generator matrix, the original choice
of $T$ matrix is $T=T_8\sqrt{3}/\sqrt{2}$ \cite{simplest} with
$T_8=\mathrm{diag}(1,1,-2)/2\sqrt{3}$ be $SU(3)$ generator. The two
$SU(3)$ triplet scalar fields in the model are expressed in terms of
pseudo Goldstone boson field $\Theta$ and their vacuum expectation
values $f_1$ and $f_2$ as \cite{simplest1}
\begin{eqnarray}
\Phi_1=e^{i\Theta\frac{f_2}{f_1}}\begin{pmatrix}0\\0\\f_1\end{pmatrix}\hspace{1cm}
\Phi_2=e^{-i\Theta\frac{f_1}{f_2}}\begin{pmatrix}0\\0\\f_2\end{pmatrix}\label{PhiDef}
\end{eqnarray}
where $f^2=f_1^2+f_2^2$,
With them the bosonic part Lagrangian of the model involving pseudo
Goldstone field  is
\begin{widetext}
\begin{eqnarray}
        \mathcal{L}&=&[ (\partial_\mu + i g A^a_\mu T^a - \frac{i}{3} g_x A^x_\mu )\Phi_i]^\dagger
        [ ( \partial_\mu + i g A^a_\mu T^a - \frac{i}{3} g_x A^x_\mu
        )\Phi_i]+\mu^2 \left(\Phi_1^\dagger \Phi_2 + \Phi_2^\dagger
\Phi_1\right)\label{Ldef}
\end{eqnarray}
\end{widetext}
with $A^a_\mu$ and $A^x_\mu$ be $SU(3)$ and $U(1)_X$ gauge fields
respectively.

 In practical calculations, for pseudo Goldstone field $\Theta$ given
 in (\ref{ThetaDef}), people use unit matrix $I$ to replace
 original $SU(3)$
 generator $T=T_8\sqrt{3}/\sqrt{2}$ \cite{simplest1,axion,axion1,decay,neutrino1}.
 The reason to replace $SU(3)$ generator  with unit matrix is explained in
 Ref.\cite{axion} in a footnote as that the choice of $SU(3)$ generator introduce kinetic
 mixing of the $\eta$ with unphysical Goldstone bosons, removing
 this  mixing by appropriate field redefinitions is equivalent to
 choosing $T$ proportional to the unit matrix. This explanation is
 not easy to read out directly from (\ref{PhiDef}) and (\ref{Ldef}).
 We have checked that the global symmetry broken is SO(6)/SO(5), both $T_8$ and
$I$ can be embedded into SO(6) and they are all belong to a same
broken generator plus some other different unbroken generators, in
this sense, two parameterizations should be identical. $I$
 commute with all the other broken generators and the mixing should be
minimal. If this judgment is correct, $T_8$ must produce same
result as that from unit matrix $I$. Consider the importance of
$\eta$ particle in the simplest little Higgs
 model, in this paper, we present a explicit proof of this equivalence by
 showing that we can choose redefined fields $\tilde{\eta}$ and $\tilde{h}$ to make the changes of two
$SU(3)$ triplet fields $\Phi_1$ and $\Phi_2$ from choice of
$T=T_8\sqrt{3}/\sqrt{2}$ to $T=I$ is equal to a local $SU(3)$
 rotation and therefore can be rotated away. i.e.
\begin{widetext}
\begin{eqnarray}
e^{id'\Theta_{\tilde{h}}}
e^{\frac{i}{f}\frac{f_2}{f_1}(\frac{\tilde{\eta}}{\sqrt{2}}I+\Theta_{\tilde{h}})}\begin{pmatrix}0\\0\\f_1\end{pmatrix}= e^{id(\frac{\sqrt{3}\eta}{2}T_8+\Theta_h)}
e^{\frac{i}{f}\frac{f_2}{f_1}(\frac{\sqrt{3}\eta}{2}T_8+\Theta_h)}
\begin{pmatrix}0\\0\\f_1\end{pmatrix}
\label{result10}
\end{eqnarray}
and
\begin{eqnarray}
e^{id'\Theta_{\tilde{h}}}
e^{-\frac{i}{f}\frac{f_1}{f_2}(\frac{\tilde{\eta}}{\sqrt{2}}I+\Theta_{\tilde{h}})}\begin{pmatrix}0\\0\\f_2\end{pmatrix}=e^{id(\frac{\sqrt{3}\eta}{2}T_8+\Theta_h)}
e^{-\frac{i}{f}\frac{f_1}{f_2}(\frac{\sqrt{3}\eta}{2}T_8+\Theta_h)}
\begin{pmatrix}0\\0\\f_2\end{pmatrix}
\label{result20}
\end{eqnarray}
\end{widetext}
where $d$ and $d'$ are two $\eta$ and $h$ dependent dimensional
parameters which make the total argument on the exponential
dimensionless, we will fix $d$ and $d'$ later. Once (\ref{result10})
and (\ref{result20}) are valid, we can take $SU(3)$ rotations to
rotate away the phase factor $e^{id'\Theta_{\tilde{h}}}$ in l.h.s.
and $e^{id(\frac{\sqrt{3}\eta}{2}T_8+\Theta_h)}$
 in r.h.s. of (\ref{result10}) and
(\ref{result20}), then choice of $T=T_8\sqrt{3}/\sqrt{2}$ and $T=I$
become equivalent. To prove
 (\ref{result10}) and
(\ref{result20}), we find they are equivalent to
 following identity
\begin{equation}
e^{i\frac{f_2}{ff_1}\frac{\tilde{\eta}}{\sqrt{2}}I+
i(d'+\frac{f_2}{ff_1})\Theta_{\tilde{h}}}\Phi_0=
e^{i(d+\frac{f_2}{ff_1})(\frac{\sqrt{3}\eta}{2}T_8+\Theta_h)}\Phi_0\label{result1}
\end{equation}
\begin{equation}
e^{-i\frac{f_1}{ff_2}\frac{\tilde{\eta}}{\sqrt{2}}I+
i(d'-\frac{f_1}{ff_2})\Theta_{\tilde{h}}}\Phi_0=
e^{i(d-\frac{f_1}{ff_2})(\frac{\sqrt{3}\eta}{2}T_8+\Theta_h)}\Phi_0\label{result2}
\end{equation}
with $\Phi_0=(0,0,1)^T$. We now discuss how to realize
(\ref{result1}) and (\ref{result2}). Because the special form the
parametrization, the exponential of $e^{i\Theta_{\tilde{h}}}$ at
l.h.s. of (\ref{result1}) and (\ref{result2}) can be work out
explicitly through explicit computations order by orders, one can
show that for some constant $c$
\begin{eqnarray}
    e^{ic\Theta_{\tilde{h}}}=I+\frac{[\cos(c\tilde{H})-1]}{\tilde{H}^2}B_{\tilde{h}}+i\frac{\sin(c\tilde{H})}{\tilde{H}}\Theta_{\tilde{h}}\label{Thetah}
\end{eqnarray}
with
$B_{\tilde{h}}=\Theta_{\tilde{h}}^2=\begin{pmatrix}\tilde{h}\tilde{h}^\dagger
    &0\\0&\tilde{H}^2\end{pmatrix}$, and $\tilde{H} = \left( \tilde{h}^\dagger \tilde{h} \right)^{\frac{1}{2}}$, Eq.(\ref{Thetah})
then leads
\begin{eqnarray}
e^{i(\frac{c'\tilde{\eta}}{\sqrt{2}}I+c\Theta_{\tilde{h}})}\Phi_0=e^{ic'\frac{\tilde{\eta}}{\sqrt{2}}}
\bigg[\cos(c\tilde{H})\Phi_0+
i\frac{\sin(c\tilde{H})}{\tilde{H}}\Phi_{\tilde{h}}\bigg]\label{result1a}
\end{eqnarray}
with definition of $\Phi_{\tilde{h}}= \begin{pmatrix}\tilde{h}\\
0\end{pmatrix}$.

Further calculations on $e^{ic(\frac{\sqrt{3}\eta}{2}T_8+\Theta_h)}$
needs some care, since $T_8$ matrix is not commute with $\Theta_h$.
Consider
\begin{eqnarray}
T_8&=&\frac{1}{2\sqrt{3}}\begin{pmatrix}1&0&0\\0&1&0\\0&0&-2\\
\end{pmatrix}=\frac{1}{2\sqrt{3}}(I-3\tilde{T})\\
\tilde{T}&=&\begin{pmatrix}0&0&0\\0&0&0\\0&0&1\\
\end{pmatrix}\nonumber
\end{eqnarray}
then we consider
\begin{eqnarray}
e^{ic(\frac{\sqrt{3}\eta}{2}T_8+\Theta_h)}\Phi_0 =
e^{i\frac{c\eta}{4}}e^{ic(-\frac{3\eta}{4}\tilde{T}+\Theta_h)}\Phi_0
\end{eqnarray}
One can check that
  \begin{equation}
  (-\frac{3\eta}{4}\tilde{T}+\Theta_h)^n =
  a_{n-1}(-\frac{3\eta}{4}\tilde{T}+\Theta_h)+a_{n-2}B_h
\end{equation}
with $a_n=(-3\eta/4) a_{n-1}+H^2 a_{n-2},~n\geq
2$,~$a_0=1$,~$a_1=-3\eta/4,~H = \left( h^\dagger h \right)^{\frac{1}{2}}$. Solve the recursion relation, we obtain
\begin{equation}
a_n=\frac{\alpha^{n+1}-\beta^{n+1}}{\alpha-\beta}
\end{equation}
where
\begin{eqnarray}
\alpha &=& \frac{-3\eta/4+ \rho}{2},\quad\beta = \frac{-3\eta/4-\rho}{2} \\
\rho &=& \sqrt{9\eta^2/16 + 4H^2}
\end{eqnarray}
With this result, then
\begin{widetext}
\begin{eqnarray}
e^{ic(-\frac{3\eta}{4}\tilde{T}+\Theta_h)}=I+[-1+e^{-\frac{3ic\eta}{8}}
(-\frac{3i\eta/4}{\rho}\sin\frac{c\rho}{2} +\cos\frac{c\rho}{2})]
\frac{B_h}{H^2}+
e^{-\frac{3ic\eta}{8}}\frac{2i\sin\frac{c\rho}{2}}{\rho}(-\frac{3\eta}{4}\tilde{T}+\Theta_h)
\end{eqnarray}
and
\begin{eqnarray}
e^{ic(-\frac{3\eta}{4}\tilde{T}+\Theta_h)}\Phi_0=e^{-\frac{3ic\eta}{8}}
[-\frac{i3\eta/4}{\rho}\sin\frac{c\rho}{2}
+\cos\frac{c\rho}{2}]\Phi_0
-e^{-\frac{3ic\eta}{8}}\frac{2i\sin\frac{c\rho}{2}}
{\rho}\Phi_h
\end{eqnarray}
Introduce  a phase angle
\begin{eqnarray}
\tan\Delta(x,y)=\frac{3x/4}{\sqrt{9x^2/16+4y^2}}\tan\frac{\sqrt{9x^2/16+4y^2}}{2}\label{DeltaDef}
\end{eqnarray}
$e^{ic(\frac{\sqrt{3}\eta}{2}T_8+\Theta_h)}\Phi_0$ then can be
expressed as
\begin{eqnarray}
e^{ic(\frac{\sqrt{3}\eta}{2}T_8+\Theta_h)}\Phi_0=e^{-
i\{\frac{c\eta}{8}+\Delta[c\eta,cH]\}}\sqrt{1-\frac{4H^2}
{\rho^2}\sin^2\frac{c\rho}{2}}~\Phi_0+
e^{-\frac{ic\eta}{8}}\frac{2i\sin\frac{c\rho}{2}}{\rho}\Phi_h\label{result2a}
\end{eqnarray}
\end{widetext}
Demand $\eta$ and $h$ fields satisfy
\begin{eqnarray}
\eta=\frac{1}{a}\tilde{\eta}\hspace{1cm}h=\frac{1}{b}~e^{-i\Delta[c\eta,cH]}
\tilde{h}\label{tildeRelation}
\end{eqnarray}
with $a$ and $b$ be two dimensionless parameters depend on $\eta$
and $h$, then compare (\ref{result2a}) with (\ref{result1a}), as
long as we demand
\begin{eqnarray}
&&b_i'a\frac{\eta}{\sqrt{2}}+\delta=-\frac{c_i\eta}{8}-\Delta[c_i\eta,c_iH]\label{ab1}\\
&&\left|\cos\left(c_i'bH\right)\right|=\sqrt{1-\frac{4H^2}{\rho^2}\sin^2\frac{c_i\rho}{2}}\label{ab2}\\
&&\frac{\sin\left(c_i'bH\right)}{H}=\frac{2\sin\frac{c_i\rho}{2}}{\rho}\label{ab3}
\end{eqnarray}
with $i=1,2$, $\delta=\pi,0$ for $\cos\left(c_i'bH\right)$ take negative and positive values
respectively, and
\begin{eqnarray}
&&c_1=d+b_1\quad c_1'=d'+b_1\quad b_1=\frac{f_2}{ff_1}\nonumber\\
&&c_2=d-b_2\quad c_2'=d'-b_2\quad b_2=\frac{f_1}{ff_2}\label{c1c2}
\end{eqnarray}
then  we will have (\ref{result1}) and (\ref{result2}). It is easy
to see (\ref{ab2}) and (\ref{ab3}) are equivalent, then only
(\ref{ab1}) and (\ref{ab3}) are independent equations from which and
(\ref{c1c2}) we can solve out parameters $a$,$b$,$d$,$d'$. With the
solution and combine with (\ref{DeltaDef}), we can rewrite
(\ref{tildeRelation}) as
\begin{eqnarray}
&&\tilde{\eta}=-\frac{c_1\eta}{4b_1'\sqrt{2}}-\frac{\sqrt{2}}{b_1'}\arctan\bigg[\frac{3\eta/4}{\rho}\tan\frac{c_1\rho}{2}\bigg]\label{tildeeta}\\
&&\tilde{h}=\frac{he^{i\Delta[c_1\eta,c_1H]}}{c_1'H}\arcsin\bigg[\frac{2H\sin\frac{c_1\rho}{2}}{\rho}\bigg]\label{tildeh}
\end{eqnarray}
in which $c_1$ and $c_1'$ are constrained by
\begin{eqnarray}
&&\frac{c_1\eta}{4b_1'\sqrt{2}}+\frac{\sqrt{2}}{b_1'}\arctan\bigg[\frac{3\eta/4}{\rho}\tan\frac{c_1\rho}{2}\bigg]\nonumber\\
&&=\frac{c_2\eta}{4b_2'\sqrt{2}}+\frac{\sqrt{2}}{b_2'}\arctan\bigg[\frac{3\eta/4}{\rho}\tan\frac{c_2\rho}{2}\bigg]~~~\\
&&\frac{1}{c_1'}e^{i\Delta[c_1\eta,c_1H]}\arcsin\bigg[\frac{2H\sin\frac{c_1\rho}{2}}{\rho}\bigg]\nonumber\\
&&=\frac{1}{c_2'}e^{i\Delta[c_2\eta,c_2H]}\arcsin\bigg[\frac{2H\sin\frac{c_2\rho}{2}}{\rho}\bigg]
\end{eqnarray}
in which from $(\ref{c1c2})$, $c_2$ and $c_2'$ related to  $c_1$ and
$c_1'$ by
\begin{eqnarray}
c_2=c_1-b_1-b_2\hspace{1cm}c_2'=c_1'-b_1-b_2\label{c1c2a}
\end{eqnarray}
Once $c_1$ and $c_1'$ is fixed, parameters $d$ and $d'$ are fixed
then through (\ref{c1c2}). Up to this stage, we achieve the proof of
(\ref{result1}) and (\ref{result2}) and then the choice of
$T=T_8\sqrt{3}/\sqrt{2}$ and $T=I$ become equivalent. This method
can be applied further to the case of that $T$ matrix is
proportional to $\tilde{T}$ and then as predicted by
Ref.\cite{axion} that this choice of $T$ matrix is also equivalent
to $T=T_8\sqrt{3}/\sqrt{2}$ and $T=I$.
\section*{Acknowledgments}

This work was  supported by National  Science Foundation of China
(NSFC) under Grant No. 10435040. We also thank Dr. David Rainwater
for helpful discussions.


\begin{thebibliography}{1}\label{biblio}

\bibitem{mHupperBound}
W.-M.Yao {\it et al.} (Particle Data Group), J. Phys. {\bf
G33},1(2006)

\bibitem{mHlowerBound}
R.Barate {\it et al.}, Phys. Lett. {\bf B565}, 61(2003)

\bibitem{littleHierarchy}
N.Arkani-Hamed, A.G.Cohen and H.Georgi, Phys. Lett. {\bf
B513},232(2001);\\
N.Arkani-Hamed, A.G.Cohen, E.Katz,A.E.Nelson, T.Gregoire and
J.G.Wacker, JHEP {\bf 0208}, 021(2002)

\bibitem{simplest}
D.E.Kaplan, M.Schmaltz, JHEP {\bf 0310},039(2003)

\bibitem{simplest1}
M.Schmaltz, JHEP {\bf 0408}, 056(2004)

\bibitem{axion}
W.Kilian, D.Rainwater and J.Reuter, Phys. Rev. {\bf
D71},015008(2005)

\bibitem{axion1}
W. Kilian, J.Reuter, D. Rainwater, hep-ph/0507081

\bibitem{decay}
K. Cheung, J.Song, hep-ph/0611294

\bibitem{neutrino}
A.Abada, G.Bhattacharyya, M.Losada,  Phys.Rev. {\bf D73},
033006(2006)

\bibitem{neutrino1}
J.Lee, hep-ph/0504136

\end{thebibliography}
\end{document}